\documentclass[%
 reprint,
 amsmath,amssymb,
 aps,
]{revtex4-1}

\usepackage{graphicx}
\usepackage{dcolumn}
\usepackage{bm}

\begin{document}

\preprint{APS/123-QED}

\title{Topological phase transformations and intrinsic size effects in ferroelectric nanoparticles}

\author{John Mangeri}
 \email{john.mangeri@uconn.edu}
\affiliation{Department of Physics, University of Connecticut}

\author{Yomery Espinal}

\affiliation{Department of Materials Science and Engineering, University of Connecticut}

\author{Andrea Jokisaari}
\affiliation{Center of Hierarchical Materials Design, Northwestern-Argonne Institute for Materials Science and Engineering, Northwestern University}

\author{S. Pamir Alpay}

\affiliation{Department of Materials Science and Engineering, University of Connecticut}
\affiliation{Department of Physics, University of Connecticut}

\author{Serge Nakhmanson}

\affiliation{Department of Materials Science and Engineering, University of Connecticut}
\affiliation{Department of Physics, University of Connecticut}

\author{Olle Heinonen}
 \email{heinonen@anl.gov}
 \affiliation{Center of Hierarchical Materials Design, Northwestern-Argonne Institute for Materials Science and Engineering, Northwestern University}
\affiliation{Material Science Division, Argonne National Laboratory}

\date{\today}

\begin{abstract}
Composite materials comprised of ferroelectric nanoparticles in a dielectric matrix 
are being actively investigated for a variety of functional properties attractive for a wide
range of novel electronic and energy harvesting devices.
However, the dependence of these functionalities on shapes, sizes, orientation and mutual arrangement 
of ferroelectric particles is currently not fully understood.
In this study, we utilize a time-dependent Ginzburg-Landau approach combined with coupled-physics
finite-element-method based simulations to elucidate the behavior of polarization in isolated spherical PbTiO$_3$ or BaTiO$_3$ 
nanoparticles embedded in a dielectric medium, including air. 
The equilibrium polarization topology is strongly affected by particle diameter, as well as the
choice of inclusion and matrix materials, with monodomain, vortex-like and multidomain patterns emerging 
for various combinations of size and materials parameters.
This leads to radically different polarization vs electric field responses, resulting in highly tunable size-dependent 
dielectric properties that should be possible to observe experimentally.
Our calculations show that there is a critical particle size below which ferroelectricity vanishes. 
For the PbTiO$_3$ particle, this size is 2 and 3.4 nm, respectively, for high- and low-permittivity media.
For the BaTiO$_3$ particle, it is $\sim$3.6 nm regardless of the medium dielectric strength.
\end{abstract}

\pacs{Valid PACS appear here}
\maketitle


\section{Introduction}
With the end of Moore's law in sight for silicon-based device technology, new paradigms utilizing alternative material families are
being probed with increasing intensity. 
Ferroelectric (FE) materials constitute one such family
that is highly attractive for a broad range of next-generation technological applications. 
FE materials are already used in non-volatile random access memories\cite{Scott1989, Auciello1998PhysicsToday, LeeAdvMat2012, ChanthboualaNature2012, GarciaNatureComm2014} 
and are now being investigated more broadly as possible components in a variety of new electronic devices, as well as for energy-storage and battery-
related technologies.\cite{Huang2010, Lichtensteiger2012, Paniagua2014}
A particularly versatile approach for infusing all of these applications with FE functionalities involves creation of composites consisting
of small FE particles dispersed within a dielectric matrix, that may be, e.g., of polymeric,\cite{Huang2010, Paniagua2014, Li2016} ferromagnetic,\cite{Etier2015} 
or oxide origin.\cite{Hu2015}
On the other hand, synthetic processes governing the formation of FE nanoparticles can produce a wide variety of shapes, including cuboidal \cite{Obrien2001, Mohanty2011, Polking2012, Caruntu2015}, ellipsoidal 
or spherical \cite{Mohanty2011, Yu2013, Qiao2015, Caruntu2015}, and core-shell\cite{Ueno2014, Huang2015, Qiao2015} geometries. 
These inclusions can be incorporated into the composite in either aggregated,\cite{Obrien2001, Huang2010} or dispersed,\cite{Caruntu2015, Li2016}
irregular arrangements, providing precise control of its dielectric properties.
However, studies of the influence of feature size on useful FE properties, and specifically the intrinsic
limit for FE response, have been largely focused on 
bulk ceramics, thin films\cite{Shaw2000,Ihlefeld2016} and bi-/multilayers,\cite{Kesim2014, Espinal2014, Maurya2015,Khassaf2016} 
with relatively few investigations dedicated to nanoparticles and other nanostructures.\cite{Akdogan2007a, Akdogan2007b, Polking2012} 
Tuning the particle size and the material parameters, as well as those of the surrounding environment, directly influences the strength of competing
energy interactions within the system, including long-range electrostatics, short-range ferroelectric ordering and electrostrictive coupling between polar 
and elastic degrees of freedom. 
With some or all of these terms being close in magnitude, the system may become 
highly sensitive to changes in control parameters, so that small external stimuli can generate large responses. 
A detailed understanding how the particle size, shape and morphology, as well as the
elastic and electrostatic influence of the surrounding medium affect FE properties is currently lacking. 
%
In this study, we focus on spherical particles, such as the ones already synthesized by a number of experimental 
groups.\cite{Mohanty2011, Yu2013, Qiao2015}
We then attempt to elucidate connections between the size of the isolated FE nanoparticle immersed in a dielectric medium (including air)
and the topological features of an equilibrium arrangement of polar dipoles within it --- what we refer to below as its \emph{polarization texture} or \emph{pattern}.
This model can be considered as equivalent to a highly dispersed particle arrangement within the matrix, where electrostatic interactions among 
individual particles are negligible.
It can also be regarded as a first step towards constructing and evaluating more complex models for composite ferroelectric-matrix systems.
In our investigation, we aim to determine specific parameter combinations, such as particle diameter and medium dielectric strength, when equilibrium 
polarization patters may become unstable. 
A topological transformation between different polarization texture morphologies can then be triggered in the vicinity of such an unstable state by an applied 
electric or elastic field, producing the desired property response.
We use two archetypical FE materials, BaTiO$_3$ (BT) and PbTiO$_3$ (PT), to represent the properties of the inclusion.
At room temperature, PT has weak electrostrictive coupling between ferroelectric polarization and elastic strain, but a large spontaneous polarization 
($P_s^\mathrm{PT} = 0.75$ C/$\mathrm{m}^2$), whereas BT has a lower polarization ($P_s^\mathrm{BT} = 0.26$ C/$\mathrm{m}^2$) but is much more 
sensitive to applied strain.
We also employ three material choices with radically different dielectric and elastic properties for the dielectric matrix: 
SrTiO$_3$ (ST, high dielectric permittivity), amorphous silica (\emph{a}-SiO$_2$, low dielectric permittivity), and vacuum (described by vacuum permittivity $\epsilon_0$).
Our findings suggest that, for a certain range of particle sizes and materials parameters, vortex-like polarization patterns are energetically favored over mono- 
or multidomain geometries. 
These patterns are not \emph{true topological} vortices and therefore a `-like' suffix is used to describe them.
Essentially similar polarization motifs have already been observed experimentally\cite{Gruverman2008, Balke2012, Gregg2012, Chae2012, Yadav2016} 
and predicted theoretically\cite{Naumov2004, Ponomareva2005, Naumov2008, Levanyuk2013, Martelli2015, Nahas2015} in some spatially confined FE 
nanostructures.
We demonstrate that particles with vortex-like and multidomain polarization patterns exhibit highly tunable multi-stage switching behavior under electric field cycling. 
Such effects should be easy to detect in experiments and may also be of use for a variety of device applications.
Additionally, we evaluate the stability of observed polarization textures depending on particle diameter, selection of materials parameters and the choice of polar 
gradient energy coefficients, which also allows us to establish a critical size for a transition into the paraelectric state. 

\section{Methods}
A general real-space finite-element approach in three dimensions (3D)  is employed to track the evolution of coupled 
polarization density, electrostatic potential and elastic displacement fields
$\mathbf{P}$, $\Phi$ and $\mathbf{u}$ in the system (boldface font marks vector fields). 
%
This approach is particularly well-suited for the studies of complex systems at mesoscale, i.e., for characteristic 
lengths that range from a few to 100s, or even 1000s of nm, which makes a uniform treatment of all the possible system sizes 
with atomistic techniques impractical.
All the numerical simulations presented here have been done with the code package \textsc{Ferret},\cite{FerretLink} which
is being developed by the authors
and is based on the Multiphysics Object-Oriented Simulation Environment (MOOSE) framework.\cite{Gaston2009}
The finite-element-based model consists of a spherical FE inclusion, $\Omega_\mathrm{FE}$, embedded in a dielectric-medium cube, 
$\Omega_\mathrm{M}$, with the whole system meshed using an unstructured grid of tetrahedrons. 
The interface between the inclusion and the matrix is assumed to be coherent. 

The following expression, adopted from thermodynamic Landau-Ginzburg-Devonshire (LGD) theory, is utilized to represent the total free energy of the system 
in the domain $\Omega_\mathrm{FE}$: 
\begin{equation}
{\mathcal F} = \! \int\limits_{\Omega_\mathrm{FE}} \! \left[f_\mathrm{bulk}+f_\mathrm{wall}+f_\mathrm{elastic}+f_\mathrm{elec}+f_\mathrm{coupled}\right] d^3\mathbf{r}.
\end{equation}
Here, $f_\mathrm{bulk}$ is the bulk ferroelectric energy density, $f_\mathrm{wall}$ is the
energy density that arises from local gradients in $\mathbf{P}$,  $f_\mathrm{elastic}$ 
is the linear elastic energy density, $f_\mathrm{coupled}$ is the
energy density due to electrostrictive coupling between the
local FE polarization density and the strain, and $f_\mathrm{elec}$ is the electrostatic energy density.  
Detailed expressions for all of the free-energy densities are provided in the Supplemental Material. 
The evolution of the polarization density field $\mathbf{P}$ is described by the time-dependent Landau-Ginzburg-Devonshire equation (TDLGD)
\begin{equation}\label{TDLGD}
- \gamma \frac{\partial \mathbf{P}}{\partial t} =  \frac{\delta}{\delta \mathbf{P}} \! \int\limits_{\Omega_\mathrm{FE}} \!\!\! d^3 \mathbf{r} \, f\!\left(\mathbf{P} \right),
\end{equation}
that drives the system towards an equilibrium state by reducing its free energy until it reaches a local minimum.
The total free energy is considered converged when its relative change between consecutive time steps is below 0.1\%.
The time constant $\gamma$ is related to polar domain-wall mobility.\cite{Meng2015}  
It is set to unity in this investigation, as we are interested not in the temporal evolution of the system, but rather only in its final (local) equilibrium state. 
As a starting guess for the $\mathbf{P}$ field, a random configuration $\langle\mathbf{P}\rangle \approx 0$, also known as a random 
paraelectric initial condition (RPEIC), is used for all the particle sizes considered in this project.
This condition allows one to avoid any initial bias or symmetry that may
restrict the path of the system during the time evolution of $\mathbf{P}$.\cite{Li2001, Li2002} 
Outside the FE inclusion, $\mathbf{P} \equiv 0$, and the behavior of the matrix is governed by the equations for the linear elastic-dielectric medium.
The coupled electrostatic potential field $\Phi$ is obtained from the Poisson equation
\begin{equation}\label{Poisson}
\nabla \cdot \left(\epsilon_\alpha \nabla \Phi \right) = - \nabla \cdot \mathbf{P}, 
\end{equation}
which, along with the condition for mechanical equilibrium,
$\nabla \cdot \sigma = 0$, has to be satisfied at each step in system time evolution governed 
by Eq.~(\ref{TDLGD}). 
This requirement implies that characteristic relaxation times for the electrostatic potential and the elastic displacement fields are much shorter
than that of the polarization-density field.
Here, $\epsilon_\alpha$ is the background dielectric constant of the FE ($\alpha=b$) that originates from polarization of the core
electrons\cite{Hlinka2006} or the dielectric constant of the matrix ($\alpha=m$), while $\sigma_{ij} = C_{ijkl} \,\partial u_k /\partial x_l$ is the stress field. 
Material parameters for the dielectric susceptibilities $\epsilon_b$ and $\epsilon_m$, and elastic coefficients $C_{ijkl}$ utilized in 
this investigation are provided in the Supplemental Material. 
%

%
As an important aside, we point out that local surface terminations of nanoparticles may be complex and 
dependent on a particular synthesis route, which could also affect properties, such as support for metal species and 
adsorption coordination.\cite{Crosby2016}
Specifically for the perovskite materials considered here, terminations consisting of $\mathrm{Pb(Ba)O}$ or 
$\mathrm{TiO}$ layers would produce different amounts of uncompensated surface charge and thus influence 
the polarization field distribution at the surface of the nanostructure.
However, by studying a wide range of dielectric constants of the surrounding matrix, $\epsilon_m$, the aggregate 
effects of charge compensation at different nanoparticle surface terminations can be effectively captured.
We also note that in cases when surface terminations may vary locally on the particle surface, 
electrostatic and elastic fields arising from such variations should rapidly average out.
The size of the cubic computational domain containing the inclusion $\Omega_\mathrm{FE}$ and the surrounding dielectric  $\Omega_\mathrm{M}$
is taken to be large enough for the elastic-displacement field $\mathbf{u}$ and the stresses $\sigma_{ij}$ arising from elastic mismatch at the interface
between the inclusion and the matrix to vanish at the domain boundaries. 
Dirichlet boundary conditions $\mathbf{u}=0$ and $\Phi=0$ are used at the $[\pm100]$, $[0\pm10]$, and $[00\pm1]$ boundaries of the computational domain. 
We can also introduce an external electrostatic field by applying a 
Dirichlet boundary condition,
$\Phi\ne0$, to the $[001]$ surface.
Consistency checks were performed to ensure that both the internal $\Phi$ and $\mathbf{u}$, in fact, do vanish at the boundaries of the computational domain
(in the absence of applied external fields) throughout the range of all the investigated inclusion diameters $d$. 
All simulations presented here are done at room temperature.
Further details of the computational method are presented in the Supplemental Material. 

\begin{figure*}[htp!] 

\center{\includegraphics[width=0.99\linewidth]{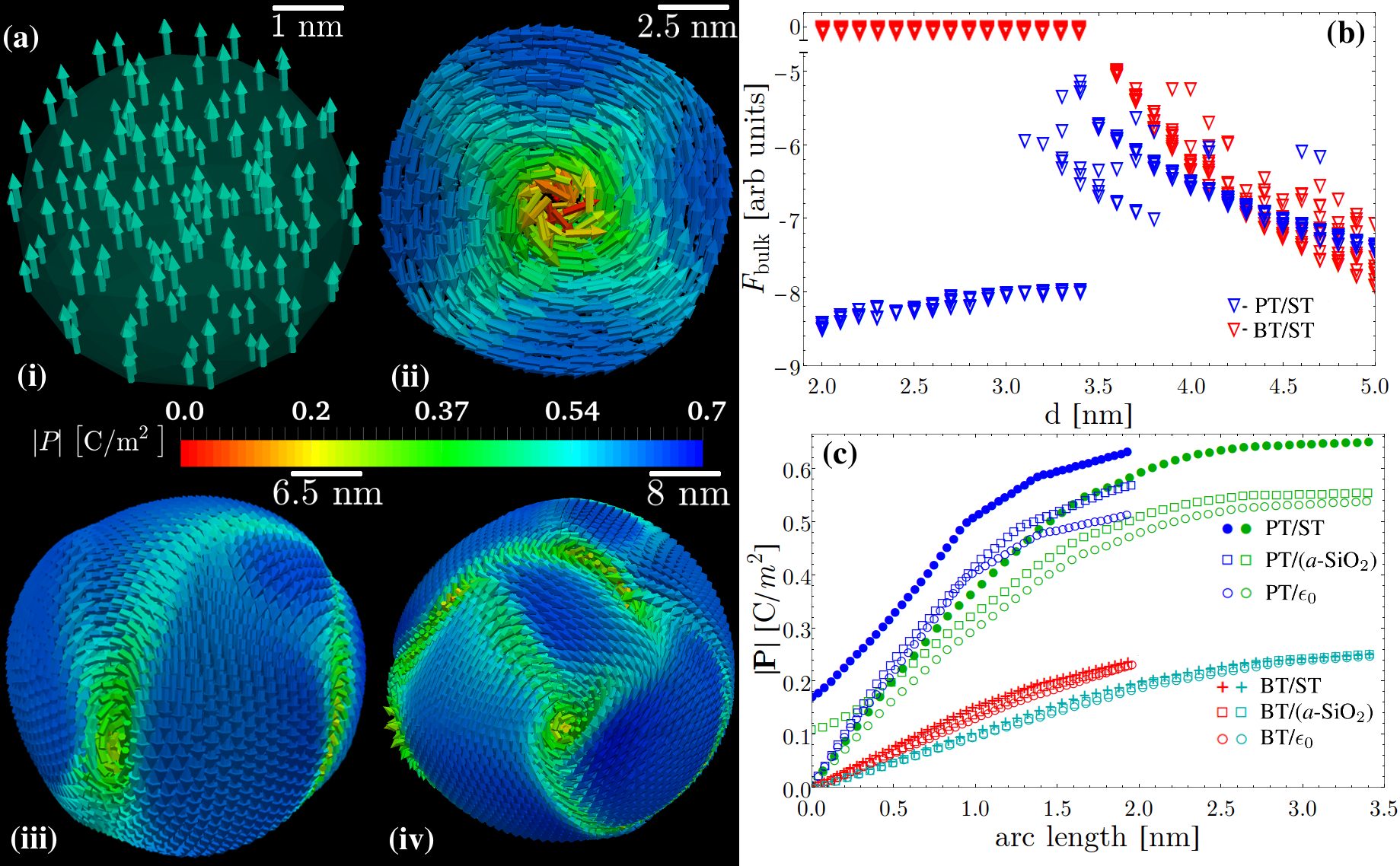}}
\caption{
(a) Various polarization-field textures observed in the PT/ST system with increasing nanoparticle diameter $d$: (i) monodomain, (ii) vortex, and (iii) - (iv) multidomain. 
Local directions of $\mathbf{P}$ are depicted by arrows, whose color represents the field magnitude.
(b) FE bulk energy $F_\mathrm{bulk}$, normalized by volume and bulk spontaneous polarization $P_s$, for PT/ST and BT/ST composites 
plotted as a function of $d$.
Monodomain-to-vortex-like phase transition occurs in the PT/ST system at critical diameter $d_v \simeq 3.4$~nm accompanied by an increase in energy. 
The BT/ST system does not form a monodomain state; instead its $\mathbf{P} \approx 0$ for $d \leq 3.6$~nm.
Above that diameter value, it transforms into a vortex-like state, which is accompanied by an energy decrease.
The reason behind the energy change during the monodomain- or paraelectric-to-vortex-like transition can be understood from panel (c), 
which shows the radial profile of $|\mathbf{P}|$ along a line perpendicular to the vortex core in 4 and 7~nm wide inclusions with
vortex-like polarization textures.
In both PT/ST and BT/ST systems, the values of $|\mathbf{P}|$ in the vortex-like phase are depressed, compared to those in the monodomain phase, or respective $P_s$.
The influence of the dielectric permittivity of the surrounding medium on the polarization values is also presented in the panel, with curves
corresponding to different dielectric materials shown for all the considered systems.
}
\label{fig:fig1}
\end{figure*}

\section{Results and discussion}
Here we discuss the topological features of polarization textures in FE nanoparticles and follow the evolution of these features with
changing particle diameter, dielectric permittivity of the surrounding matrix and applied electric field.
For all of the considered systems, we can identify two distinct transitions as the diameter of the inclusion increases: from
a paraelectric or polar-monodomain to a vortex-like state, and from a vortex-like to a polydomain state 
(see Fig.~\ref{fig:fig1}(a) for examples).
While these transitions are common to the materials systems investigated, their details, as well as the 
particle sizes at which they occur, depend on the specific materials parameters.

\subsection{Paraelectric and ferroelectric states for small $d$}
The disappearance of the system FE polarization below a certain critical diameter $d_c$ is
found to be strongly dependent on the selection of materials parameters for the inclusion and the dielectric matrix.
In general, i.e., for both of the FE inclusion materials considered here, we obtain values of $d_c$, above which a non-zero polarization 
distribution $\mathbf{P}$ may exist in some form, as spanning from 2 to $\sim$ 3.6~nm.
These critical lengths fall within an approximate range identified by other research groups for a variety of different FE structures, 
including thin films.\cite{Fong2004, Erdem2006, Polking2012, Grunebohm2012, Ihlefeld2016} 
They are also substantially smaller than the FE correlation length $\ell_\mathrm{C}$\cite{Glinchuk2008, Levanyuk2013} ---
i.e., a typical size for supporting a single FE domain wall --- which for these nanoparticles we estimate as being $\sim$ 5--10~nm.
Simulations conducted for systems with particles having $d \lessapprox d_c$
provided the following insights into the specifics of their evolution towards an equilibrium configuration.
At the beginning of the simulation, the $f_\mathrm{wall}$ energy term is large due to inhomogeneities in the $\mathbf{P}$
distribution produced by the RPEIC. 
In a small nanoparticle with $d \ll \ell_\mathrm{C}$, $\mathbf{P}$ initially evolves towards a monodomain state, which 
produces a non-zero surface charge density $q_S=\mathbf{P}\cdot\hat n$, where $\hat n$ is the surface normal vector. 
In turn, the presence of uncompensated $q_S$ leads to a sharp increase in the $f_\mathrm{elec}$ energy term,
unless the dielectric permittivity of the matrix is large enough to screen out the surface charge.
In order to reduce $f_\mathrm{elec}$, the magnitude of $\mathbf{P}$ is uniformly diminished until $\textbf{P} \approx 0$
within numerical precision of the simulation, resulting in the paraelectric state of the system.
We verified that the robustness of the paraelectric solution for $d<d_c$ does not depend on the initial conditions specifics 
or finite-element mesh spacing if the interfacial region between the inclusion and the matrix is sufficiently resolved.\cite{Li2001, Li2002} 
For a PT inclusion, a monodomain FE state is observed for $d<$~3.4~nm for $\epsilon_m \gtrapprox 300$ (i.e., ST matrix),
with a uniform distribution of local polarization. 
The polarization magnitude is somewhat reduced, compared to $P_s^\mathrm{PT}$, as shown in sketch (i) of Fig.~\ref{fig:fig1}(a).  
For lower values of $\epsilon_m$, a paraelectric state is found for the PT inclusion, while in the BT system such state persists for all the values 
of $\epsilon_m$, including one corresponding to the ST matrix.
That may be surprising, considering that $P_s^\mathrm{BT} \ll P_s^\mathrm{PT}$,
which must reduce $q_S$ and therefore reduce the penalty originating from the $f_\mathrm{elec}$ energy term for the monodomain state of the BT system. 
However, due to the shallower bulk energy minimum and stronger electrostrictive couplings in the BT system, the energy increases 
arising from the $f_\mathrm{bulk}$ and $f_\mathrm{coupled}$ terms are relatively minor, and thus 
the system evolution towards the equilibrium is still dominated by the influence of the $f_\mathrm{elec}$ term and minimization of $q_S$.
Therefore, for all of the investigated particle/matrix material combinations except PT/ST, the first FE state with non-zero local $\mathbf{P}$ that is encountered
as the particle size is increased is the one that has a vortex-like character.
\subsection{Transition into a vortex-like state}

The inception of the vortex-like state in both PT and BT nanoparticles can be detected by following the dependence of their normalized bulk free energy, 
\begin{equation}
F_\mathrm{bulk} = \frac{1}{\Omega_\mathrm{FE} P_s^3} \int\limits_{\Omega_\mathrm{FE}} \! f_\mathrm{bulk}\,d^3\mathbf{r}, 
\end{equation}
on $d$, as shown in Fig.~\ref{fig:fig1}(b) for the ST dielectric matrix. 
For the PT/ST system, the transition occurs at a critical diameter $d_v\simeq$~3.4~nm and is accompanied by a sharp \emph{increase} of $F_\mathrm{bulk}$,
compared to its value in the monodomain state. 
On the other hand, in the BT/ST system, $F_\mathrm{bulk}$, which is zero by definition in the paraelectric state, sharply \emph{decreases} 
upon the formation of the vortex-like state with non-zero polarization at $d_v \equiv d_c \simeq$~3.5~nm. 
While the physics underpinning the energy change in the BT/ST system may be self-evident, understanding the behavior of the PT/ST system requires a 
detailed examination and comparison of the polarization patterns before and after the transition, i.e., sketches (i) and (ii) depicted in Fig.~\ref{fig:fig1}(a).
As can be seen from sketch (ii), even though local values of $|\mathbf{P}| \sim P_s^\mathrm{PT}$ 
near the surface of the inclusion, they become strongly suppressed close to its center, forming a weakly polar or even completely paraelectric core region of
the vortex. 
In all of the PT systems, this core region is cylindrical in shape and penetrates the spherical inclusion completely from its northern to southern pole, as
shown in Supplemental Fig.~1(a).
In the BT systems, such a region is also present, but its shape may be twisted or bent, as illustrated in Supplemental Fig.~1(b), and its final conformation 
exhibits dependence on the RPEIC. 
The change in the value of $|\mathbf{P}|$ along the direction perpendicular to the vortex core axis is presented in Fig.~\ref{fig:fig1}(c)
for particles of two different sizes for both PT and BT systems coupled with all the considered dielectric matrices.
These data show that $|\mathbf{P}|$
in the core region may be reduced by a factor of 3--5, compared to its surface value 
(in PT), or even disappear completely (in PT and always in BT).
This tendency is mostly unaffected by the dielectric strength of the surrounding matrix.
Such behavior is in sharp contrast with that of ferromagnetic vortices, where, at temperatures well below $T_\mathrm{C}$, magnetization density at the core is 
constrained to a constant magnitude.\cite{Lee2014,Zhou2015} 
Fig.~\ref{fig:fig1}(c) also describes the effect of the surrounding matrix on the value of $|\mathbf{P}|$ at the surface of the inclusion.
In the PT system, surface polarization is $\sim20\%$ larger when it is coupled with a high dielectric permittivity medium, such as ST.
In contrast, surface polarization of the BT system is not affected by the strength of the dielectric screening provided by the matrix.   
Topological features of polarization textures, summarized in Fig.~\ref{fig:fig1}, are in general agreement with the
analytical work of Levanyuk and Blinc.\cite{Levanyuk2013} 
The observed similarities include the dependence of $|\mathbf{P}|$ on the surrounding medium, its 
suppression at the core of the vortex-like phase, as well as transitory nature of monodomain states in low dielectric permittivity medium.  
However, the investigation of Levanyuk and Blinc did not consider coupling between ferroelectric and elastic degrees of freedom.
Although the results presented here include the effects of electrostrictive coupling, we have also performed a series of simulations with 
the electrostrictive tensor set to zero in order to examine its influence on the behavior of the system.
%
%
The resulting polarization textures have sharp 90$^\circ$ domain walls, resembling Landau flux-closure 
patterns found in some magnetic microstructures.\cite{SchaferBook, Raabe2005} 
Thus, we conclude that the presence of the electrostrictive coupling is responsible for the softening 
of the domain walls, which produces more rounded textures, such as the ones that were observed or predicted in other experimental and 
theoretical studies.\cite{Naumov2008, Gruverman2008, Nahas2015} 
Representative images of polarization textures formed with and without the electrostrictive coupling in a PT/ST system are
shown in Supplemental Fig.~2.
\begin{figure}[htp!] 
\center{\includegraphics[width=0.99\linewidth]{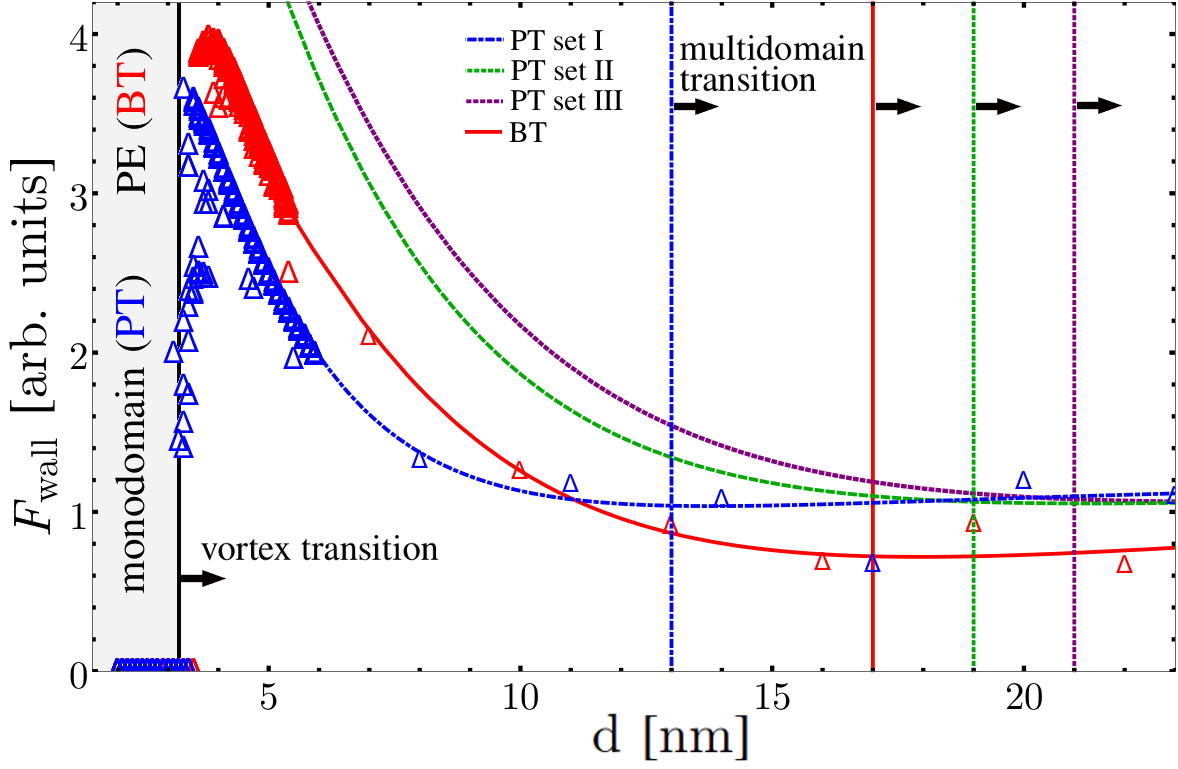}}
\caption{\small{
Normalized gradient energy $F_\mathrm{wall}$\cite{FitFootnote} as a function of $d$ for the PT/ST and BT/ST systems.
$F_\mathrm{wall}$ is zero for $d < d_v$, but increases rapidly in the vortex-like state because of sub-optimal arrangements of the
local polarization vectors. 
It then levels off, i.e., becomes bulk-like, in the multidomain state, with the transition point (marked by vertical lines) depending rather sensitively 
on the choice of $G_{ijkl}$ parameters for PT.
}}
\label{fig:fig2}
\end{figure}

\subsection{Subsequent transition into multidomain state}

\begin{figure*}
\centering
\includegraphics[width=1\linewidth]{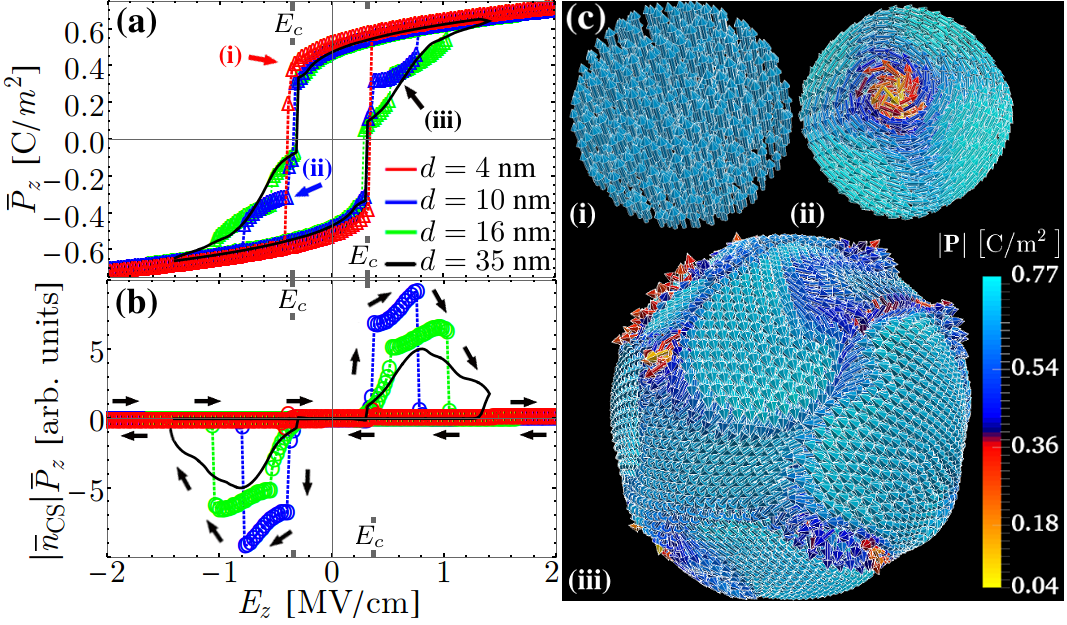}
\caption{\label{fig:PT_hyst}
(a) Average polarization $\bar{P}_z$ and (b) polarization-scaled Chern-Simons number density $| \bar{n}_\mathrm{CS} | \cdot \bar{P}_z$ 
as functions of applied electric field $E_z$ in PT/ST structures of different sizes.
Panel (c) shows polarization textures in (i) monodomain, $d = 4$~nm, (ii) vortex-like, $d = 10$~nm, and (iii) multidomain, $d = 35$~nm, structures
corresponding to the markings in panel (a).
}
\end{figure*}

For particle diameters, $d > d_v$, the most important energy term influence further polarization texture evolution is the normalized gradient energy,
\begin{equation}
F_\mathrm{wall} = \frac{1}{\Omega_\mathrm{FE} P_s^3} \int\limits_{\Omega_\mathrm{FE}} \! f_\mathrm{wall}\,d^3\mathbf{r}. 
\end{equation}
In the case of BT, the well-known parameterization of Hlinka and Marton\cite{Hlinka2006} can be used for the gradient energy tensor $G_{ijkl}$.
However, a number of various parameterizations for $G_{ijkl}$ exist for PT.
In this investigation, we considered three different sets --- attributed to Li \emph{et al.}\cite{Li2001, Li2002}\ (set I), Wang \emph{et al.}\cite{Wang2004}\ (set II) and
Hong \emph{et al.}\cite{Hong2011}\ (set III).
The values of the $G_{ijkl}$ coefficients used in all of these sets, as well as for BT, are listed in the Supplemental Material.
The dependence of $F_\mathrm{wall}$ on $d$ is shown in Fig.~\ref{fig:fig2} for PT/ST and BT/ST systems.
The gradient energy is zero in monodomain and paraelectric states for $d < d_v$.
It then grows rapidly upon transition to the vortex-like state, which can be construed as consisting of a large number 
of domain walls separating small polar regions that have energetically sub-optimal mutual polarization arrangements 
(as opposed to optimal ones of $90^\circ$ and $180^\circ$).
As the inclusion diameter increases beyond $d_v$, $F_\mathrm{wall}$ gradually recedes until it saturates at a constant non-zero value.
Such leveling off indicates the formation of a `bulk-like' multidomain state comprised of relatively large areas of 
correlated \textbf{P} divided by domain walls that are similar to their $90^\circ$ and $180^\circ$ bulk variants.
This transition happens at $d \equiv d_m \simeq$~17~nm for the BT/ST system, while in PT/ST its arrival is quite sensitive to the choice of the
$G_{ijkl}$-coefficient set and covers the range of 13 to 21 nm.
However, the equilibrium topologies of multidomain states obtained for PT/ST do not seem to be strongly affected by the choice of $G_{ijkl}$
parameterizations.
We \emph{speculate} that by measuring certain experimentally observable quantities linked to this transition --- e.g., 
electric field response that is discussed next ---  it may be possible to obtain better gradient energy estimates for PT.
Representative images of multidomain polarization textures are shown as cases (iii -- iv) in Fig.~\ref{fig:fig1}(a),
as well as case (iii) in Fig.~\ref{fig:PT_hyst}(c).
We find that domains always tend to orient their polarization tangentially to the surface of the inclusion in order to minimize the 
electrostatic energy arising from $q_S$.
The non-polar or weakly polar vortex core area becomes unstable at $d > d_m$, developing uniform polarization and eventually
splitting into multiple domains.
The apparent vorticity of the polarization texture sharply decreases after the transition into the multidomain state, but it does not 
disappear completely, as localized vortices, marked by largely suppressed $\mathbf{P}$, still remain near some domain walls.  
Due to the curvature of the inclusion surface, the observed domain patterns cannot be directly partitioned into
collections of low-energy $90^\circ$ or $180^\circ$ variants in the near-surface region, which, in combination with remaining vorticity, 
results in finite saturation values of $F_\mathrm{wall}$ at large $d$. 
Quantitative evaluations of the amount of vorticity present in a three-dimensional vector field can be conducted by
computing of its Chern-Simons topological winding number density\cite{GeraldDunneLectures} $n_\mathrm{CS}.$
In the case of $\textbf{P},$ $n_\mathrm{CS}  =  \left(\nabla \times \textbf{P}\right) \cdot \textbf{P}.$
%
%
It should be close to zero for monodomain and bulk-like multidomain patterns, and must have an extremum
in the vortex-like phase that separates them, as shown in Supplemental Fig.~3.
We find this measure to be especially useful in elucidating the topological changes of polarization textures in inclusions subjected to
applied electric fields, as discussed below.
\subsection{Intrinsic field dependence of polarization}
We have studied the electric-field induced topological changes for all of the polarization textures --- monodomain, vortex-like and multidomain ---
that we observe in both PT and BT-based FE nanoparticles, thus probing their intrinsic response to applied fields. 
Such simulations are done by first applying an external electric field $E_z \equiv - E_\mathrm{max}\hat z$ to volume $\Omega_\mathrm{M}$, which induces
a saturated monodomain configuration in the particle with the polarization aligned with the field.
After the saturated configuration is established, the field is increased in small steps to $+E_\mathrm{max}\hat z$ and then decreased 
back to $-E_\mathrm{max}\hat z$, thus completing the poling loop. 
At each step, the averaged projection on the Cartesian $\hat z$-axis, $\bar{P_z},$ is computed for the converged polarization field $\mathbf{P}$.
Note that initial poling curves starting from zero applied field and zero polarization are not simulated with this approach.
For PT, we use the $G_{ijkl}$ parameterization set I\cite{Li2001, Li2002} in all calculations.
Further details of the poling method implementation are provided in the Supplemental Material. 
In Fig.~\ref{fig:PT_hyst}(a) we show field induced polarization response in PT/ST structures of four different sizes, exhibiting 
monodomain [case (i), $d = 4$~nm], vortex-like [case (ii), $d = 10$~nm] and multidomain [$d = 16$~nm and case (iii), $d = 35$~nm] 
zero-field equilibrium polarization textures, some of which are also visualized in panel (c) of the same figure.
In the monodomain case, the tetragonal crystallographic axis, or the ``easy'' polarization axis, is aligned with the $\hat z$ direction. 
The associated poling loop is similar to that of a generic bulk FE, with abrupt switching of $\mathbf{P}$ between 
$-\hat z$ and $+\hat z$ orientations at a distinct value of coercive field $E_c$.
A comparison of field induced polarization responses from different crystallographic PT orientations is shown in Supplemental Fig.~4.
For particle diameters $d > d_v$, switching between $-\hat z$ and $+\hat z$ monodomain orientations proceeds in two stages.
The initial monodomain configuration persists from $E_z = -E_\mathrm{max}$ to $E_z \simeq + E_c$, at which point it
is replaced with a hybrid texture consisting of a monodomain core polarized along the $+\hat z$ direction combined with a vortex-like 
closure pattern in the near-surface region.
This results in the response curve exhibiting a small plateau at $E_c < E_z < E_\mathrm{max}$ with roughly constant $\bar{P}_z$, 
originating from the polarized inclusion core, that is considerably smaller than the saturation polarization.  
As the field is increased further, the vortex-like texture is abruptly expelled from the near-surface region and the 
polarization aligns along the $+\hat z$ direction everywhere in the particle.
Multistage switching processes similar to the one observed here have been reported before in some nanopatterned FE\cite{Gruverman2008, Martelli2015} 
and ferromagnetic\cite{Boardman2004, Boardman2005} systems. 
As the particle diameter increases further, the intermediate vortex-like texture occurring in the the near-surface region at $E_z \simeq + E_c$ 
is replaced with a multidomain texture, such as the one shown as case (iii) in Fig.~\ref{fig:PT_hyst}(c).
Similar to the zero-field configurations, domains at the inclusion surface prefer to have tangential orientations of their $\mathbf{P}$ to minimize 
the electrostatic energy arising from $q_S$.
As the magnitude of the field is increased,
the transition into the monodomain state occurs by a gradual alignment of the surface domain polarizations along the $+\hat z$ direction.
This switching mechanism produces multiple shoulders in the response curve, which merge together smoothly for larger particles that contain 
many surface domains [see, e.g., curve (iii) in Fig.~\ref{fig:PT_hyst}(a)].

\begin{figure*}
\centering
\includegraphics[width=0.98\linewidth]{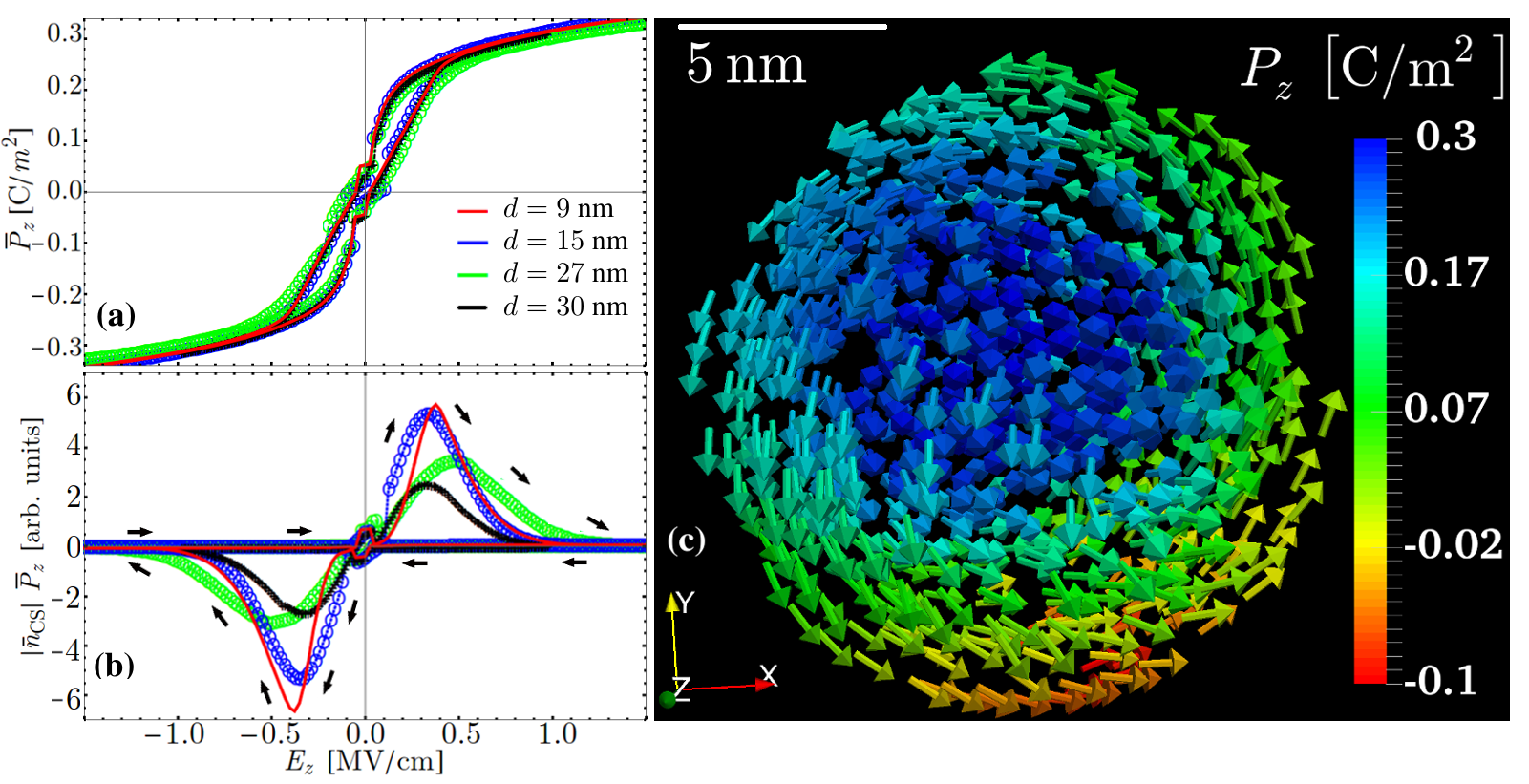}
\caption{\label{fig:BT_hyst}
(a) Average polarization $\bar{P}_z$ and (b) polarization-scaled Chern-Simons number density $\left| \bar{n}_\mathrm{CS} \right| \cdot \bar{P}_z$ 
as functions of applied electric field $E_z$ in BT/ST structures of different sizes.
Panel (c) shows a BT/ST system with a 15~nm inclusion that is in the process of nucleating a monodomain core (the blue region) as its polarization 
aligns with the external field. 
At zero applied field, this particle has a vortex-like $\mathbf{P}$ texture. 
In the near-surface region, $\mathbf{P}$ is curling around, which results in nonzero value of $n_\mathrm{CS}$.
}

\end{figure*} 

In Fig.~\ref{fig:PT_hyst}(b) we present the changes in system vorticity under the changing electric field, which is represented by the 
sphere-averaged value of  $\left| \bar{n}_\mathrm{CS} \right|$ weighted by the $\bar{P}_z$, for the same PT/ST structures as in Fig.~\ref{fig:PT_hyst}(a).
In case (i), monodomain texture at zero field, polarization switching occurs with $\left| \bar{n}_\mathrm{CS} \right| \equiv 0$ everywhere throughout the
poling loop.
For all the other cases at $d > d_v$, switching between monodomain states at $-E_\mathrm{max}\hat z$ and $+E_\mathrm{max}\hat z$
happens through the formation of an intermediate vortex-like state, as indicated by the non-zero value of averaged $n_\mathrm{CS}$. 
In case (ii), vortex-like texture at zero field, vorticity changes abruptly, but, as can be seen from comparison of the curves for $d = 16$ and 35~nm,
this transition becomes progressively more diffuse for zero-field multidomain textures at increasing $d$.
However, even for large nanoparticles containing many domains, such as in case (iii), $\left| \bar{n}_\mathrm{CS} \right|$ remains non-zero 
during the switching due to the presence of vortex-like twists of $\mathbf{P}$ along domain walls.
Fig.~\ref{fig:BT_hyst} presents field induced variations in polarization response [panel (a)] and vorticity [panel (b)] in the
BT/ST systems with particle $d = 9$ to 30~nm.
These curves look more slim, in comparison with the ones shown for PT/ST in Fig.~\ref{fig:PT_hyst}(a),
while the dependence of $\left| \bar{n}_\mathrm{CS} \right|$ on $E_z$ suggests diffuse poling behavior in BT/ST proceeding through the formation
of an intermediate state with non-zero vorticity at all of the considered inclusion sizes.
Therefore, unlike in the PT/ST systems with $d_v < d < d_m$, where vortex-like texture in the near-surface region gets expelled abruptly
upon transitioning into the monodomain state [see curve (ii) in Fig.~\ref{fig:PT_hyst}(a-b)], in BT/ST this texture disappears
gradually, as local polarization continuously rotates to align itself with the applied field and the core monodomain grows outward.  
A vector map of such an intermediate hybrid state, exhibiting both monodomain and vortex-like features, is shown in Fig.~\ref{fig:BT_hyst}(c). 
All these results suggest that a wide variety of different switching patterns and behaviors can be designed by controlling the size of the particle 
as well as the materials properties of the particle and the matrix. 
We note that the dielectric response of an aggregate system consisting of FE particles of varying sizes dispersed in a dielectric medium 
will depend strongly on the nature of the applied mechanical and electrical boundary conditions. 
For example, coherency (misfit) strains between the particle and the dielectric matrix, thermal and/or epitaxial stresses introduced into thin-film 
heterostructures during growth, specifics of the electric field application to the structure --- such as usage of top-bottom or interdigitated electrodes, 
or an atomic-force microscope tip --- will all have a significant effect on the overall dielectric response of the composite. 
\section{Conclusions}
We have investigated the behavior of ferroelectric nanoparticles in dielectric media in a parametric space where different interactions, 
such as the electrostatics resulting from surface and bulk charges, as well as domain-wall and electrostrictive energies, are of similar 
magnitudes and compete.
As a consequence, the observed ferroelectric behavior and response is complex and highly tunable by the selection of materials parameters
and/or external fields. 
Our results show that a high-permittivity dielectric medium, that compensates charges on the inclusion surface, can stabilize 
non-zero polarization in PT particles as small as 2 nm in diameter. 
In contrast, embedding in a low-permittivity medium results in large uncompensated surface charges that completely suppress polarization 
in particles smaller than $d \simeq$ 3.4--3.6~nm. 
Above that critical size, the FE state emerges as a vortex-like texture, which minimizes the
electrostatic energy arising from the surface charges, while at even larger sizes, a multidomain texture is formed as a compromise between 
the electrostatic and polarization-gradient energy contributions. 
The electrostrictive coupling between the elastic and polar degrees of freedom softens sharp $90^\circ$ domain walls, producing
rounded vortex-like textures with cylindrical cores that can penetrate all the way through the spherical particle.
From an intricate dependence of the shape of the field vs polarization response loop on the particle size --- an effect that cannot
be directly reproduced in bulk FE materials --- we also predict high intrinsic dielectric tunability of such FE nanoinclusions.
This behavior is rooted in a multistage switching of the polarization through an intermediate state with non-zero Chern-Simons vorticity that can 
emerge/disperse gradually or abruptly, depending on a particular choice of material parameters and particle sizes.
Field induced polarization response curves, such as the ones presented in Figs.~\ref{fig:PT_hyst}(a) and \ref{fig:BT_hyst}(a), can be obtained 
experimentally, e.g., using piezo-force microscopy.
Therefore, it may be possible to utilize such measurements to explore the predicted rearrangements of polarization textures within the FE particles, 
as well as to evaluate the quality of the LGD $G_{ijkl}$-parameterizations for the gradient energy terms by observing and comparing the shapes of 
the response curves in samples with different particle sizes.
\section{Acknowledgments}
The authors are indebted to Dmitry Karpeyev for significant contributions to the \textsc{Ferret} repository. 
J.M. acknowledges funding support from the U.S. Department of Energy, Office of Science, Office of Workforce Development for Teachers and Scientists, 
Office of Science Graduate Student Research (SCGSR) program.
The SCGSR program is administered by the Oak Ridge Institute for Science and Education (ORISE) for the DOE.
ORISE is managed by ORAU under contract number DE-SC0014664.
The work by O.H. was funded by the US Department of Energy, Office of Science, Basic Energy Sciences, Division of Materials Science and Engineering.
The work of A.M.J. was performed under financial assistance award 70NANB14H012 from U.S. Department of Commerce, 
National Institute of Standards and Technology as part of the Center for Hierarchical Material Design (CHiMaD). 
J.M. would also like to thank Candost Akkaya for a helpful discussion.
The authors also acknowledge the computing resource support provided on Blues, a high-performance computing cluster operated by the 
Laboratory Computing Resource Center at Argonne National Laboratory, and on the Hornet cluster, hosted by the Taylor L.\ Booth 
Engineering Center for Advanced Technology, located at the University of Connecticut at Storrs.

\bibliography{rsc} 

\begin{mcitethebibliography}{59}
\bibitem[Fit()]{FitFootnote}
Fit functions are of the form $f(d) = c_1 \exp{\left(-c_2 d\right)} + c_3/d +
  c_4$. We find that, likely within error, the fit parameter $c_3$ strongly
  depends on $G_{110}$. On this graph, after $d > 5$ nm, only every fifth data
  point is shown for BT and PT set I.\relax
\mciteBstWouldAddEndPunctfalse
\mciteSetBstMidEndSepPunct{\mcitedefaultmidpunct}
{}{\mcitedefaultseppunct}\relax
\EndOfBibitem
\end{mcitethebibliography}
\bibliographystyle{rsc} 
\end{document}